\pdfoutput=1

\documentclass[11pt,twoside,a4paper,cmspaper,final,collab]{cms-tdr}
\def\svnVersion{494065P}\def\cmsCernNoTag{CERN-EP-2019-001}\def\cmsCernDate{\today}\def\cmsMessage{Published in Physical Review Letters as \href{http://dx.doi.org/10.1103/PhysRevLett.122.151802}{\doi{10.1103/PhysRevLett.122.151802.}}}

\begin{document}\cmsNoteHeader{SMP-18-009}

\hyphenation{had-ron-i-za-tion}
\hyphenation{cal-or-i-me-ter}
\hyphenation{de-vices}
\RCS$HeadURL: svn+ssh://svn.cern.ch/reps/tdr2/papers/SMP-18-009/trunk/SMP-18-009.tex $
\RCS$Id: SMP-18-009.tex 491675 2019-03-13 13:11:34Z ccaillol $
\newlength\cmsFigWidth
\ifthenelse{\boolean{cms@external}}{\setlength\cmsFigWidth{0.49\textwidth}}{\setlength\cmsFigWidth{0.65\textwidth}}
\ifthenelse{\boolean{cms@external}}{\providecommand{\cmsLeft}{upper\xspace}}{\providecommand{\cmsLeft}{left\xspace}}
\ifthenelse{\boolean{cms@external}}{\providecommand{\cmsRight}{lower\xspace}}{\providecommand{\cmsRight}{right\xspace}}
\ifthenelse{\boolean{cms@external}}{\providecommand{\CL}{C.L.\xspace}}{\providecommand{\CL}{CL\xspace}}
\newcommand{\mppp}{\ensuremath{m_{3\Pgp}}\xspace}

\cmsNoteHeader{SMP-18-009}
\title{Search for \texorpdfstring{$\PW$}{W} boson decays to three charged pions}

\date{\today}

\abstract{
For the first time, a search for the rare decay of the $\PW$ boson to three charged pions has been performed. Proton-proton collision data recorded by the CMS experiment at a center-of-mass energy of 13\TeV, corresponding to an integrated luminosity of 77.3\fbinv, have been analyzed. No significant excess is observed above the background expectation. An upper limit of $1.01\times10^{-6}$ is set at 95\% confidence level on the branching fraction of the $\PW$ boson to three charged pions. This provides a strong motivation for theoretical calculations of this branching fraction.
}

\hypersetup{
pdfauthor={CMS Collaboration},
pdftitle={Search for W boson decays to three charged pions},
pdfsubject={CMS},
pdfkeywords={CMS, physics, standard model, rare W decays}}

\maketitle

Exclusive decays of the $\PW$ boson to hadrons, if observed, would provide a new precision measurement of the mass of the $\PW$ boson that is based solely on visible decay products.
This as yet unobserved probe of the strong interaction at the boundary of perturbative and nonperturbative domains could also provide
insight into quantum chromodynamics (QCD), in particular factorization and meson form factors at high energy scales~\cite{Grossmann:2015lea,Melia:2263121,Perez:2015lra}.
Upper limits on the branching fractions ($\mathcal{B}$) of two exclusive $\PW$ boson decays were set previously at a 95\% confidence level (\CL):
$\mathcal{B}(\PW^\pm\to\PD_\text{s}^\pm\cPgg)<1.3\times 10^{-3}$~\cite{Abe:1998vm} and $\mathcal{B}(\PW^\pm\to\Pgp^\pm\cPgg)<7.0\times10^{-6}$~\cite{Aaltonen:2011nn}.
For the $\PZ$ boson, there are many more results on searches for such rare exclusive decays~\cite{PhysRevD.98.030001}.
The decay $\PW^\pm\to\Pgp^\pm\Pgp^\pm\Pgp^\mp$ could be the first exclusive decay of the $\PW$ boson to hadrons to be observed because the low particle multiplicity in the final state provides a clean signature in a collider experiment. There exists no theoretical calculation of $\mathcal{B}(\PW^\pm\to\Pgp^\pm\Pgp^\pm\Pgp^\mp)$, but it is expected to be the same order of magnitude as $\mathcal{B}(\PZ\to\Pgp^+\Pgp^-\Pgp^0)$, and within the range $10^{-8}$ and $10^{-5}$~\cite{Melia:2263121,Mangano:2014xta}.

This Letter reports a first search for the rare $\PW^\pm\to\Pgp^\pm\Pgp^\pm\Pgp^\mp$ decay. The analysis relies on an innovative use of algorithms designed to trigger on and identify $\Pgt\to\text{hadrons}+\nu_\Pgt$ weak decays~\cite{TAU-14-001,Sirunyan:2018pgf}.
The data correspond to an integrated luminosity of 77.3\fbinv recorded in proton-proton ($\Pp\Pp$) collisions at a center-of-mass energy of 13\TeV in 2016 (35.9\fbinv) and 2017 (41.4\fbinv). The two data-taking periods are analyzed separately and then combined. For simplicity, in the following the charges of the particles are omitted.

The central feature of the CMS apparatus is a superconducting solenoid 6\unit{m} in
internal diameter, providing a magnetic field of 3.8\unit{T}. A silicon pixel and strip tracker, a lead tungstate crystal
electromagnetic calorimeter (ECAL), and a brass and scintillator hadron calorimeter reside within the solenoid volume.
Each of these systems is composed of a barrel and two endcap sections.
Forward hadron calorimeters extend
the pseudorapidity ($\eta$) coverage provided by the barrel and endcap detectors. Muons are
detected in gas-ionization chambers embedded in the steel flux-return yoke
outside the solenoid. Events of interest are selected using a two-tiered trigger
system~\cite{Khachatryan:2016bia}.
A more detailed description of the CMS detector, together with a definition of
its coordinate system and relevant kinematic variables, can be found
in Ref.~\cite{Chatrchyan:2008zzk}.

The signal process $\PW\to 3\Pgp$ is simulated at leading order (LO) in perturbative QCD using the Monte Carlo (MC) event generator
\PYTHIA 8.212~\cite{Sjostrand:2014zea}. The transverse momentum spectrum of the $\PW$ boson ($\pt(\PW)$ spectrum) is corrected to match that obtained in a $\PW\to\ell\nu$ ($\ell=\Pe,\Pgm,\Pgt$) simulation in \MGvATNLO v2.3.3~\cite{Alwall:2014hca} using the jet matching and merging scheme simulated at next-to-LO (NLO) through the FxFx algorithm~\cite{Frederix:2012ps}.
The $\PZ/\cPgg^{(*)}\to\ell\ell$ background is generated at LO using \MGvATNLO with the MLM jet matching algorithm~\cite{Alwall:2007fs}.
Small background contributions from $\PW\PW$, $\PW\PZ$, and $\PZ\PZ$ (diboson) production are simulated at NLO with
\MGvATNLO, whereas the $\ttbar$ process is simulated with the \POWHEG
2.0~\cite{Nason:2004rx,Frixione:2007vw, Alioli:2008tz, Alioli:2010xd, Alioli:2010xa}
generator.
The generators are interfaced to \PYTHIA
to model parton showering and fragmentation.
The \PYTHIA parameters affecting the description of the underlying event are set to the {CUETP8M1} tune~\cite{Khachatryan:2015pea}.
The NNPDF 3.0~\cite{Ball:2014uwa} parton distribution functions (PDF), with the order matching that used in the MC generation, are used in the simulation of all processes.
Generated events are processed through a simulation of the CMS detector based on
\GEANTfour~\cite{Agostinelli:2002hh}, and are reconstructed with the same algorithms used for data.
Additional $\Pp\Pp$ collisions in the same or nearby bunch crossings (pileup) are simulated, and reweighted to reproduce the pileup multiplicity measured in data.

The reconstruction of events relies on the particle-flow (PF) algorithm~\cite{Sirunyan:2017ulk}, which combines information from the subdetectors to reconstruct and identify individual particles in an event.
Combinations of these PF objects are used to reconstruct
higher-level objects such as jets and hadronically decaying $\tau$ leptons, denoted as $\tauh$.
The reconstructed vertex with the largest value of summed physics-object $\pt^2$ is the
primary $\Pp\Pp$ interaction vertex. The physics objects are the jets, clustered using a jet-finding
algorithm~\cite{Cacciari:2008gp,Cacciari:2011ma} with the tracks assigned to the vertex as inputs, and the associated
missing transverse momentum is the negative vector sum of the \pt of those jets.
The jets are reconstructed from PF objects with the anti-\kt algorithm~\cite{Cacciari:2008gp} implemented
in the \FASTJET library~\cite{Cacciari:2011ma, Cacciari:fastjet2}, using a distance parameter of 0.4.

Charged pions are reconstructed using the hadrons-plus-strips (HPS) algorithm~\cite{TAU-14-001,Sirunyan:2018pgf}, which is primarily designed to reconstruct $\tauh$ candidates.
This allows us to align the offline object reconstruction with the requirements of the analysis trigger, which are based on the presence of two high-\pt $\tauh$ candidates,
and to use advanced discriminators developed for the HPS algorithm to reject quark and gluon jets, electrons, and muons.
The HPS algorithm is seeded using jets, and combines tracks and energy depositions in the ECAL to form $\tauh$ candidates.
Three decay modes are reconstructed: 1-prong, 1-prong + $\Pgp^0$, and 3-prong~\cite{TAU-14-001,Sirunyan:2018pgf}.
The 1-prong decay mode of the HPS algorithm targets $\Pgt$ leptons decaying to a charged pion and a $\nu_\Pgt$.
About 85\% of $\Pgt$ leptons decaying to a charged pion and a $\nu_\Pgt$, and 15\% of $\Pgt$ leptons decaying to a charged pion, a neutral pion, and a $\nu_\Pgt$, are reconstructed as 1-prong $\tauh$ candidates. 
Charged-pion candidates are required to be reconstructed in the 1-prong decay mode of the algorithm.
They need to be isolated, in that the sum of the \pt of charged particles and photons with $\pt>0.5\GeV$ reconstructed with the PF algorithm around the $\tauh$ object within a cone of $\Delta R  = \sqrt{\smash[b]{(\Delta \eta)^2 + (\Delta \phi)^2}} = 0.3$ is required to be less than 2\GeV, where $\phi$ denotes the azimuthal angle measured in radians. This isolation criterion has an efficiency above 90\% for each pion.
The pion candidates must also pass veto requirements for electrons and muons~\cite{TAU-14-001,Sirunyan:2018pgf}.
The discriminator against muons makes use of information from the muon detectors. The efficiency for true $\tauh$ or pions is larger than 99\%,
while the misidentification rate is about $3.5\times10^{-3}$~\cite{Sirunyan:2018pgf}. The discriminator against electrons uses various hadron calorimeter and ECAL variables. Its efficiency is about 80\%, with the misidentification rate of the order of $3\times 10^{-3}$~\cite{Sirunyan:2018pgf}. Pions must meet the minimum requirement
that the distance of closest approach to the primary vertex satisfies $\abs{d_z}<0.01\unit{cm}$
along the beam direction, and $d_{xy}<0.01\unit{cm}$ in the transverse plane.

Events are selected using triggers that require the presence of two $\tauh$ candidates with \pt thresholds greater than
35 or 40\GeV, depending on the run. About 3\% of the signal events have two pions with $\pt>35\GeV$ at generator level.
Offline, the two pion candidates that correspond to the trigger objects are required to be reconstructed in the 1-prong $\tauh$ decay mode, and to have $\pt>35\GeV$ (or 40\GeV, depending on the trigger)
and $\abs{\eta}<2.1$. In addition, a third pion candidate is required to be present in the event, and be reconstructed as a 1-prong $\tauh$ candidate with $\pt>18\GeV$ and $\abs{\eta}<2.3$.
The selected 1-prong $\tauh$ candidates are required to be separated from each other by at least $\Delta R  = 0.3$, be isolated, and pass the discriminator against muons. They must not all have the same electric charge. At least two of the pions are also required to pass the discriminator against electrons.
The reconstructed \pt of the $\PW$ boson, computed as the vectorial sum of the $\vec{\pt}$ of the
three pion candidates, is required to be greater than 40\GeV. This selection criterion has an efficiency greater than 90\% for signal events passing the other criteria listed above,
because the $\PW$ bosons must have sufficiently high \pt to pass the trigger requirements, and rejects more than two thirds of the
background events.

The background is dominated by standard model events composed of jets produced through the strong interaction, referred to as QCD multijet events.
This component is estimated from data, using a control region (CR1) with events that contain at least one HPS-reconstructed pion
candidate failing the isolation condition.
Such events are reweighted using an extrapolation factor that depends on the invariant mass of the three pions, $\mppp$, to obtain the QCD multijet
contribution in the signal region (SR). The extrapolation factor is obtained from events complementary to the SR, selected with $\pt(\PW)<40\GeV$. The dependence on $\mppp$ is parameterized with a linear function. Because of the poor statistical uncertainty in the region with
 $\pt(\PW)<40\GeV$, the chosen linear dependence on mass suffices to parameterize the $\mppp$ spectrum. The contribution 
to CR1 of processes other than QCD multijet events, in particular $\PZ/\cPgg^{(*)}\to\ell\ell$ events, is estimated from simulations and subtracted. 

The second largest background corresponds to $\PZ/\cPgg^{(*)}\to\ell\ell$ events. The contribution from $\PZ/\cPgg^{(*)}\to\Pgt\Pgt$ events, with $\Pgt$ decays to one charged hadron and a neutrino, is dominant, but $\PZ/\cPgg^{(*)}\to\Pe\Pe$ and 
$\PZ/\cPgg^{(*)}\to\Pgm\Pgm$ events also enter the SR because electrons and muons can easily be wrongly reconstructed as isolated 1-prong $\tauh$ candidates by the HPS algorithm.
The normalization of the $\PZ/\cPgg^{(*)}\to\ell\ell$ background
is estimated from simulation using the cross section computed in \FEWZ 3.1.b2~\cite{Li:2012wna}.
The \pt of the $\PZ$ boson and the mass spectra found in the LO simulation are corrected to match the data using correction factors obtained from $\PZ\to\Pgm\Pgm$ events~\cite{Sirunyan:2017khh}.
The $\mppp$ distribution for  $\PZ/\cPgg^{(*)}\to\ell\ell$ background in the SR comes from another control region in data (CR2),
similar to the SR, but with the exception that two of the pion candidates are required to fail the discriminator that rejects electrons misidentified as $\tauh$ candidates. The CR2 is enriched in $\PZ/\cPgg^{(*)}\to\Pe\Pe$ events, but the $\mppp$ distribution has been verified to be compatible for all $\PZ/\cPgg^{(*)}\to\ell\ell$ decays.

The yields of the diboson, single top quark, $\ttbar\PZ$, $\ttbar\PW$, and triboson backgrounds are small, and these processes are estimated from simulation.
The normalization of the $\ttbar$ background is estimated from simulation, while the $\mppp$ distribution for these events is taken from data in a different control region (CR3) where there is at least one jet identified as originating from a bottom quark~\cite{Sirunyan:2017ezt}, and where the pion candidates pass relaxed criteria that discriminate against muons and electrons.

The results are obtained by using a binned maximum likelihood fit to the $\mppp$ distribution. The mass resolution for the signal process is about 7\%. Systematic uncertainties are treated as nuisance parameters in the fit. The statistical uncertainty of the collected data constitutes about 90\% of the total uncertainty in the final results.

The uncertainty in the integrated luminosity, applied to the signal and to backgrounds estimated from simulation, amounts to 2.5\% in 2016~\cite{CMS-PAS-LUM-17-001} and 2.3\% in 2017~\cite{CMS-PAS-LUM-17-004}.
Because of the limited size of the simulated samples in the SR, and data in CR1, CR2, and CR3,
statistical uncertainties in individual bins of the $\mppp$ distributions are treated
as Poissonian nuisance parameters for all processes.

The leading systematic uncertainty is related to the pion identification efficiency and amounts to 5\% per pion. This corresponds to an uncertainty of 15\% for an event with 3 pions. The magnitude of the uncertainty is determined through a dedicated measurement in $\PZ\to\Pgt_\Pgm\tauh$ events, where the $\tauh$ candidates are reconstructed in the 1-prong decay mode and $\Pgt_\Pgm$ denotes a $\Pgt$ lepton decaying to a muon and neutrinos.
The uncertainty in the trigger efficiency is uncorrelated with the pion identification efficiency uncertainty and amounts to an additional 3.5\% for each pion accepted by the trigger~\cite{Sirunyan:2018pgf}.
For pions with $35<\pt<40\GeV$, the trigger efficiency uncertainty is inflated by 10\% because this \pt region corresponds to a turn-on of the trigger, where uncertainties in the energy scale of trigger objects lead to large uncertainties in the efficiency.
The signal sample contains about 65\% of the events that have both triggering pions with $\pt>40\GeV$, and 30\% that have one triggering pion
with $35<\pt<40\GeV$.
The uncertainty in the pion energy scale is also measured in $\PZ\to\Pgt_\Pgm\tauh$ events, and amounts to 1.2\%~\cite{Sirunyan:2018pgf}.
It affects both the normalization and the $\mppp$ distribution for simulated processes with pions
or genuine $\tauh$ leptons.
All these uncertainties are uncorrelated between the 2016 and 2017 data.

The uncertainty in the QCD multijet background comes from two sources. The first source reflects the parameterization of the extrapolation factor as a function of $\mppp$ for events with at least one nonisolated pion candidate in CR1.
The uncertainty in the slope of the linear function is propagated to the $\mppp$ spectrum and treated as an uncertainty in the dependence of the distribution.
The second source is related to the background estimation method, which makes use of events where at least one pion candidate is not isolated.
This uncertainty amounts to 30\%, and is also uncorrelated between the two data-taking periods.
The magnitude of the uncertainty is estimated from the results of three studies.
The first compares the default background estimate to that obtained from reweighting events where one of the pions fails the isolation condition but still
has an isolation value below a weaker threshold as high as 6\GeV,
resulting in a normalization difference up to 18\%. In the second study,
the default background estimate is compared to that obtained when the extrapolation factor is parameterized as a function of the leading pion \pt or of $\pt(\PW)$. These alternative parameterizations lead to variations up to
20\% in the QCD multijet background normalization. And in a third study, the agreement between data and predicted backgrounds is verified
in a dedicated control region (CR4), where at least one of the pion candidates fails the $d_{xy}$ or $\abs{d_{z}}$ requirements.
The difference between the predicted backgrounds and data amounts to 23\% in CR4.
The uncertainty related to the subtraction of simulated events in CR1 is negligible with respect to the statistical uncertainty in the $\mppp$ distribution of the QCD multijet background.

The theoretical uncertainty in the yield of the $\PZ/\cPgg^{(*)}\to\ell\ell$ background is 2\%, as determined in \FEWZ 3.1.b2~\cite{Li:2012wna}.
The number of simulated $\PZ/\cPgg^{(*)}\to\ell\ell$ events selected in the SR is small, and a statistical uncertainty in the normalization must be included explicitly because a smoothed $\mppp$ distribution from CR2 is used in the SR. The statistical uncertainty in the $\PZ/\cPgg^{(*)}\to\ell\ell$ yield amounts
to 30 (20)\% in 2016 (2017).
The cross sections of the diboson production processes and their uncertainties arise from next-to-NLO computations made using \textsc{MCFM} (v8.0)~\cite{Campbell:2010ff}.
The full next-to-NLO with next-to-next-to-leading logarithmic order resummation of soft gluon corrections~\cite{Beneke:2011mq,Cacciari:2011hy,Baernreuther:2012ws,Czakon:2012pz,Czakon:2012zr,Czakon:2013goa},
performed with the \textsc{Top++} 2.0 program~\cite{Czakon:2011xx}, is used to compute a \ttbar production cross section equal to $832^{+40}_{-45}\unit{pb}$ using a top quark mass of 172.5\GeV.

The theoretical uncertainty in the yield of $\PW$ bosons is 4\%, as determined with \FEWZ 3.1.b2~\cite{Gavin:2012sy,Melnikov:2006kv}. It includes PDF and scale uncertainties.
An uncertainty in the reweighting of the \PYTHIA signal sample to NLO as a function of the generated $\pt(\PW)$ is also included.
It amounts to 5\% and is dominated by statistical uncertainties. As verified in $\PZ\to\Pgm\Pgm$ events, the difference in the $\pt$ spectra of the vector bosons between data and NLO simulation is at the subpercent level, and is therefore neglected.

\begin{figure}[!htb]
  \centering
    \includegraphics[width=0.49\textwidth]{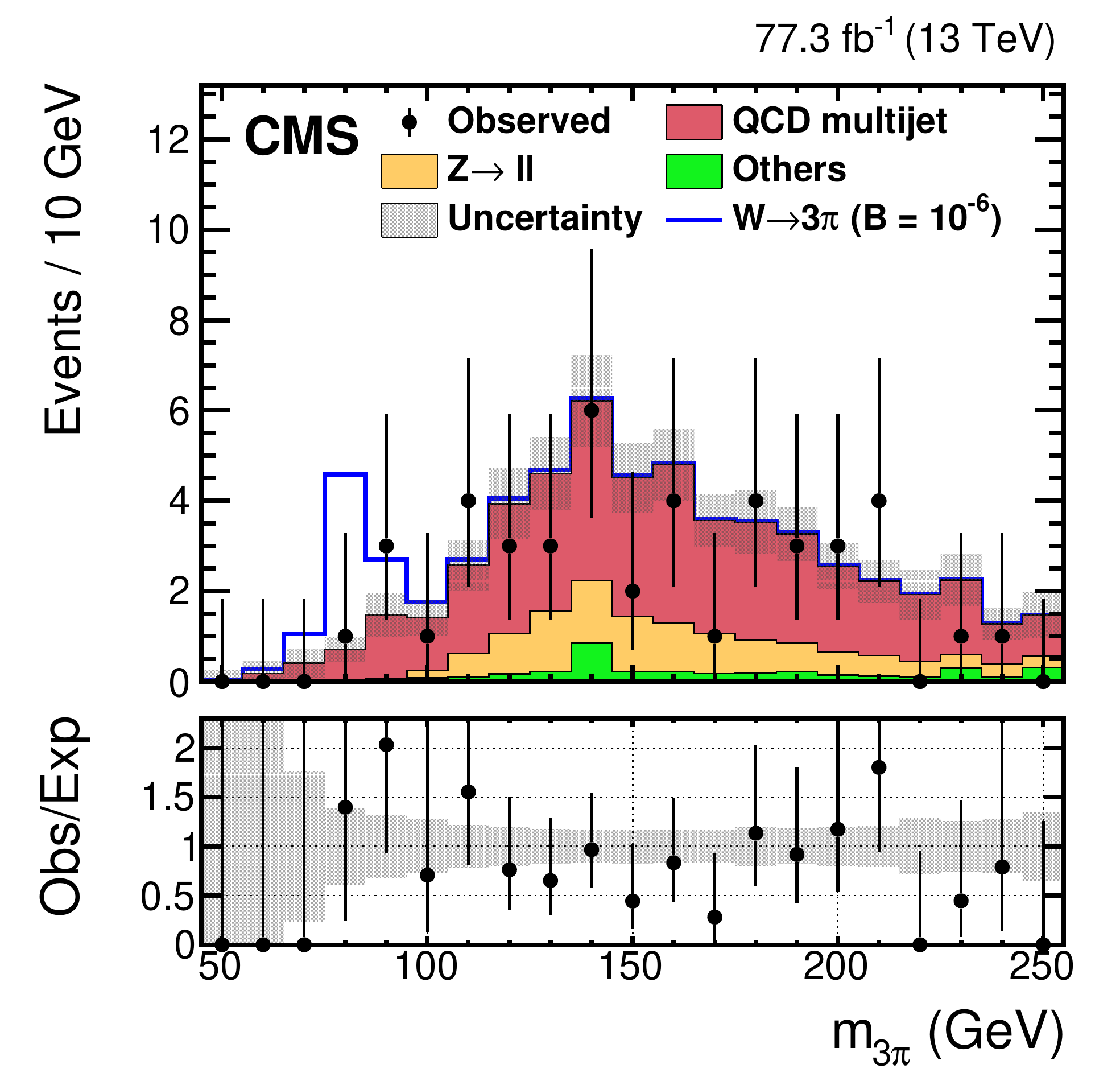}
   \caption{Observed and expected distributions of $\mppp$, using 77.3\fbinv of data collected in 2016 and 2017.
The normalization of the expected backgrounds results from the maximum likelihood fit to the data, which includes all systematic uncertainties.
 The background contributions are stacked upon each other, and the ``Others" background contribution includes events from diboson, triboson,
$\ttbar$, single top quark, $\ttbar\PZ$, and $\ttbar\PW$ production.
The uncertainty band includes both statistical and systematic sources of uncertainty.
The signal is normalized to $\mathcal{B}(\PW\to 3\Pgp)=10^{-6}$, and is shown as an open stacked histogram.
}
    \label{fig:SR}
\end{figure}

\begin{figure}[!htb]
\centering
    \includegraphics[width=0.49\textwidth]{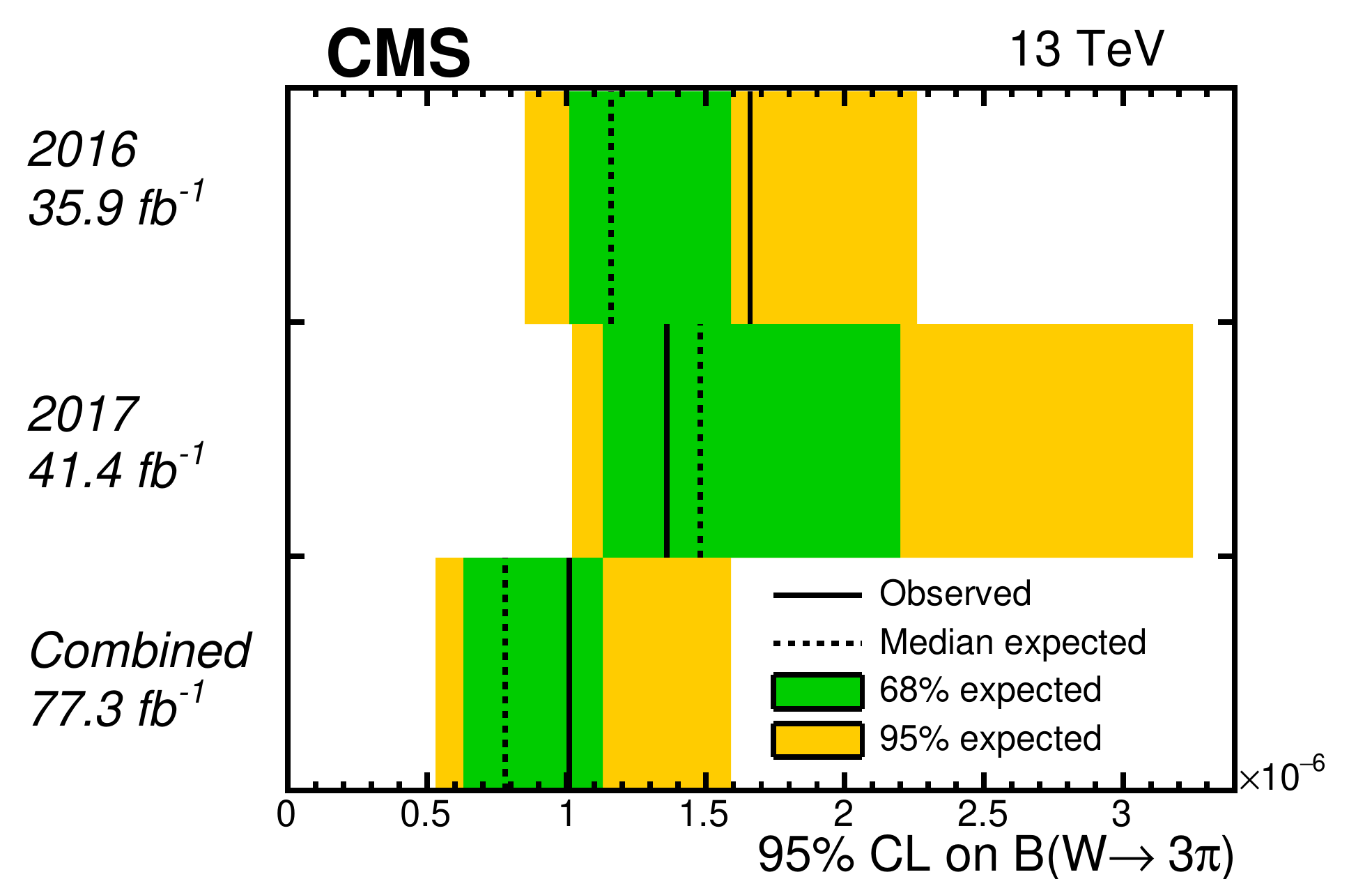}
   \caption{Observed and expected upper limits on $\mathcal{B}(\PW\to 3\Pgp)$ at 95\% \CL. The expected limit corresponds to the assumption $\mathcal{B}(\PW\to 3\Pgp)=0$. }
    \label{fig:limit}
\end{figure}

Combining both data-taking periods, the $\mppp$ distribution is shown in Fig.~\ref{fig:SR}.
No significant excess above the expected background is observed.
Upper limits at a 95\% \CL are set on the branching fraction of the $\PW$ boson to three charged pions using the \CLs method~\cite{LHC-HCG-Report,Junk,Read2} and are shown in Fig.~\ref{fig:limit}. These are $1.66\times 10^{-6}$ in the 2016 data, $1.36\times 10^{-6}$ in the 2017 data. The combination of the two data sets leads to:
\begin{displaymath}
\mathcal{B}(\PW\to 3\Pgp)<1.01\times10^{-6}.
\end{displaymath}
The expected sensitivity using the 2017 data is slightly worse than using the 2016 data despite the larger integrated luminosity because of higher \pt thresholds to trigger the events during part of the run.

In summary, the first search for the rare decay of the $\PW$ boson to three charged pions has been performed. The data correspond to an integrated luminosity of 77.3\fbinv of proton-proton collisions collected in the CMS detector at a center-of-mass energy of 13\TeV. Events with $\PW$ bosons are selected with a trigger designed to identify events with two $\Pgt$ leptons decaying to hadrons and a neutrino. The data are compatible with the background hypothesis. The upper limit on the branching fraction of the $\PW$ boson to three charged pions is $1.01\times10^{-6}$ at 95\% \CL, which reduces the previous range of values and motivates theoretical calculations of this branching fraction.

\begin{acknowledgments}
We congratulate our colleagues in the CERN accelerator departments for the excellent performance of the LHC and thank the technical and administrative staffs at CERN and at other CMS institutes for their contributions to the success of the CMS effort. In addition, we gratefully acknowledge the computing centers and personnel of the Worldwide LHC Computing Grid for delivering so effectively the computing infrastructure essential to our analyses. Finally, we acknowledge the enduring support for the construction and operation of the LHC and the CMS detector provided by the following funding agencies: BMBWF and FWF (Austria); FNRS and FWO (Belgium); CNPq, CAPES, FAPERJ, FAPERGS, and FAPESP (Brazil); MES (Bulgaria); CERN; CAS, MoST, and NSFC (China); COLCIENCIAS (Colombia); MSES and CSF (Croatia); RPF (Cyprus); SENESCYT (Ecuador); MoER, ERC IUT, and ERDF (Estonia); Academy of Finland, MEC, and HIP (Finland); CEA and CNRS/IN2P3 (France); BMBF, DFG, and HGF (Germany); GSRT (Greece); NKFIA (Hungary); DAE and DST (India); IPM (Iran); SFI (Ireland); INFN (Italy); MSIP and NRF (Republic of Korea); MES (Latvia); LAS (Lithuania); MOE and UM (Malaysia); BUAP, CINVESTAV, CONACYT, LNS, SEP, and UASLP-FAI (Mexico); MOS (Montenegro); MBIE (New Zealand); PAEC (Pakistan); MSHE and NSC (Poland); FCT (Portugal); JINR (Dubna); MON, RosAtom, RAS, RFBR, and NRC KI (Russia); MESTD (Serbia); SEIDI, CPAN, PCTI, and FEDER (Spain); MOSTR (Sri Lanka); Swiss Funding Agencies (Switzerland); MST (Taipei); ThEPCenter, IPST, STAR, and NSTDA (Thailand); TUBITAK and TAEK (Turkey); NASU and SFFR (Ukraine); STFC (United Kingdom); DOE and NSF (USA).
\end{acknowledgments}

\bibliography{auto_generated}

\cleardoublepage \appendix\section{The CMS Collaboration \label{app:collab}}\begin{sloppypar}\hyphenpenalty=5000\widowpenalty=500\clubpenalty=5000\vskip\cmsinstskip
\textbf{Yerevan Physics Institute, Yerevan, Armenia}\\*[0pt]
A.M.~Sirunyan, A.~Tumasyan
\vskip\cmsinstskip
\textbf{Institut f\"{u}r Hochenergiephysik, Wien, Austria}\\*[0pt]
W.~Adam, F.~Ambrogi, E.~Asilar, T.~Bergauer, J.~Brandstetter, M.~Dragicevic, J.~Er\"{o}, A.~Escalante~Del~Valle, M.~Flechl, R.~Fr\"{u}hwirth\cmsAuthorMark{1}, V.M.~Ghete, J.~Hrubec, M.~Jeitler\cmsAuthorMark{1}, N.~Krammer, I.~Kr\"{a}tschmer, D.~Liko, T.~Madlener, I.~Mikulec, N.~Rad, H.~Rohringer, J.~Schieck\cmsAuthorMark{1}, R.~Sch\"{o}fbeck, M.~Spanring, D.~Spitzbart, W.~Waltenberger, J.~Wittmann, C.-E.~Wulz\cmsAuthorMark{1}, M.~Zarucki
% ...
\vskip\cmsinstskip
% ...
1:  Also at Vienna University of Technology, Vienna, Austria\\
% ...
\end{sloppypar}
\end{document}